\documentclass[12pt]{article}

\usepackage{amsmath,amssymb,epsfig}
\usepackage{graphicx}

\begin{document}
\def\refname{\Large~~~~~~~{\bf References}}
\newcommand{\el}{\left}
\newcommand{\er}{\right}
\newcommand{\p}{\prime}
\newcommand{\rr}{\rho}
\newcommand{\ro}{\rho^\circ}
\newcommand{\ti}{\tilde}
\newcommand{\veps}{\varepsilon}
\newcommand{\dis}{\displaystyle}
\newcommand{\scr}{\scriptsize}
%\large
%\vspace{1mm}

\begin{center}
{\Large \bf
Restoration of Heavy-Ion Potentials at Intermediate Energies
%and the Problem of Their Ambiguity
}\\[5mm]
{\large\bf  ~K.M.~Hanna$^{1}$, K.V~Lukyanov$^2$, V.K.~Lukyanov$^2$,
B.~S{\l}owi{\'n}ski$^{3,4}$,
E.V.~Zemlyanaya$^2$}\\[3mm]
{\small\it ~~~~~$^1$Math. and Theor. Phys. Dept., NRC, Atomic Energy
Authority,
Cairo, Egypt}\\
{\small\it $^2$Joint Institute for Nuclear Research, Dubna,
Russia~~~~~~~~~~~~~~~~~~~~~~~~~~~~}\\
{\small\it $^3$Faculty of Physics, Warsaw University of
Technology, Warsaw, Poland ~~~}\\
{\small\it $^4$Institute of Atomic Energy, Otwock-Swierk,
Poland~~~~~~~~~~~~~~~~~~~~~~~~~~~~}\\
\vspace{0.5cm}
\end{center}
%\vspace{0.2cm}

\vspace{0.2cm}

{\small
The microscopic nucleus-nucleus optical potential is constructed basing
on two patterns for real and imaginary parts, each calculated in the
framework of microscopic models and multiplied by two normalizing
factors, the free parameters, fitted to experimental data. The first
supplementary model yields the real and imaginary templates for our
potential, and itself reproduces the scattering amplitude of the microscopic
Glauber-Sitenko theory. The other pattern, for real part only, is the
standard double-folding model with the exchange term included. As a result,
we obtain an acceptable agreement with elastic differential cross-sections.
}

\vspace{0.5cm}

\def\baselinestrech{1.5}

{\normalsize
%\large
\section {Introduction}

In the preceding paper \cite{K1} we have developed a method for restoration
of nucleus-nucleus potentials basing on the Glauber-Sitenko microscopic
theory of scattering at high energies \cite{{Gla},{Sit}}, generalized in
\cite{{Czyz},{Form}} to the nucleus-nucleus scattering. As a result this
theory derives the microscopic eikonal phase of scattering without
introducing an optical potential. In the so-called optical limit, this
phase is determined by the point nucleon density distributions
of the projectile(target) nucleus, $\ro_p(r)$($\ro_t(r)$), and of
the nucleon-nucleon scattering amplitude, and can be expressed as follows
%[1]
\begin{equation}\label{eq1}
\Phi_N(b)={\bar{\sigma}_{NN}\over 2}(i+{\bar\alpha}_{NN})
\int d^2s_p\, d^2s_t~\ro_p(s_p)~\ro_t(s_t)~f_N(|\bf\xi=\bf b+{\bf s}_p-
{\bf s}_t|),
\end{equation}
where $\ro(s)=\int_{-\infty}^\infty \ro(\sqrt{s^2+z^2})dz$ is the profile
function of $\ro(r)$ and
%[2]
\begin{equation}\label{eq2}
f_N(\xi)=(2\pi)^{-2}\int d^2q~\exp(-i{\bf q}{\bf\xi}) {\ti f}_N(q).
\end{equation}
Here ${\ti f}_N(q)$ is the form factor of the NN scattering amplitude, that
is usually taken in the Gaussian shape ${\ti f}_N(q)=\exp(-q^2r_{N\,rms}^2/6)$
with $r_{N\,rms}$, the NN interaction ${rms}$ radius. Here $\bar\sigma_{NN}$
is the total cross section of the NN scattering while $\bar\alpha_{NN}$ is
the ratio of the real-to-imaginary part of the forward NN scattering
amplitude, both depended on energy. We denote that the "bar" means averaging
on isotopic spins of colliding nuclei.\\
The idea of \cite{K1} consists in comparison of the microscopic phase
(\ref{eq1}) with phenomenological one defined through the optical potential
$U(r)=V(r)\,+\,iW(r)$ as follows
%[3]
\begin{equation}\label{eq3}
\Phi(b) = -\frac 1 {\hbar v}\, \int_{-\infty}^\infty U\el(\sqrt{b^2+z^2}\er)dz,
\end{equation}
where $v$ is the relative motion velocity. When doing so the analytic
expression has been used of the phase (\ref{eq3}), obtained in \cite{LZ}
for the symmetrized Woods-Saxon potential, which is the mostly realistic
phenomenological potential applied in many calculations. In \cite{K1}
parameters of this potential were adjusted so that to fit the shape of the
phenomenological phase to the microscopic one in the outer region of space
$b\sim R_p\,+ \,R_t$. As a result of this procedure we obtained a set of
the SWS-potentials having the same tails but different interiors, and
nevertheless all of them led to an acceptable agreement with the same
elastic differential cross-section. Thus, utilizing this method, one turns
out in the face of the certain problem of an ambiguity of the restored
potentials.

%\large
\section {The microscopic potentials}

In these circumstances, at the present paper we introduce the other approach to
restore potential. We believe that the microscopic models of potentials
give us a more reliable basis to search realistic potentials of scattering,
than fitting phenomenological forms of potentials.

 As a first candidate towards the search of the realistic
optical potential we take the potential which unambiguously corresponds to
the microscopic phase (\ref{eq1}) of the in high-energy approximation (HEA).
Indeed, it has been shown that this potential can be obtained when applying
the inverse Fourier transform to the HEA phase (\ref{eq1}) \cite{S1} or,
independently, in \cite{lzl}, by substituting the standard expression for
the direct double folding potential \cite{SL} in the definition of phase
(\ref{eq3}). As a result one gets the so-called HEA optical potential:
%[4]
\begin{equation}\label{eq4}
U^H_{opt}(r)\,=\, V^H(r)\,+\,iW^H(r),
\end{equation}
%[5]
\begin{equation}\label{eq5}
V^H(r)=-{2E\over k(2\pi)^2}\sigma_{NN}\alpha_{NN}
\int dq~q^2j_0(qr){\ti\ro}_p(q){\ti\ro}_t(q){\ti f}_N(q),
\end{equation}
%[6]
\begin{equation}\label{eq6}
W^H(r)=-{2E\over k(2\pi)^2}\sigma_{NN}
\int dq~q^2j_0(qr){\ti\ro}_p(q){\ti\ro}_t(q){\ti f}_N(q).
\end{equation}
Here ${\ti\ro}_{p(t)}(q)$ are form factors of the corresponding
point densities $\ro_{p(t)}(r)$ of nuclei, and the latter can be
obtained by unfolding the nuclear densities $\rr_{p(t)}(r)$ (see,
e.g.,\cite{lzs}), which are usually reported in tables. Thus, the
model does not use free parameters when calculating $V^H$ and
$W^H$ potentials. The important and novel point is that it
provides to calculate the imaginary potential microscopically
(\ref{eq6}). Indeed, in the standard semi-microscopic model one
estimates only the real potential with a help of the double
folding (DF) procedure, while the imaginary part is usually taken
in a phenomenological Woods-Saxon (WS) form with three or more
fitted parameters. In our method, we also apply this model for the
real part of an optical potential. As a matter of fact the
DF-model includes both the direct and exchange terms of the
potential (see, e.g., \cite{Knyaz}, \cite{KS})
%[7]
\begin{equation}\label{eq7}
V^{DF}=V^D~+~V^{EX}
\end{equation}
%[7a]
$$
\qquad V^D(r) = \int d^3 r_p d^3 r_t \, \rho_p({\bf r}_p)\,
\rho_t({\bf r}_t)\, v_{NN}^D({\bf r}_{pt}), \quad
{\bf r}_{pt}={\bf r}+{\bf r}_t-{\bf r}_p,
\eqno (7a)
$$
%[7b]
$$
\qquad V^{EX}(r) = \int d^3 r_p d^3 r_t \, \rho_p({\bf r}_p, {\bf r}_p+
{\bf r}_{pt})\,\rho_t({\bf r}_t, {\bf r}_t-{\bf r}_{pt})\times \hspace*{4cm}
$$
$$
\hspace*{8cm}  v^{EX}_{NN}({\bf r}_{pt})\,\exp\el[{i{\bf K}(r)
{\bf r}_{pt}\over M}\er],
\eqno (7b)
$$
Here the main dependence on energy of the potential occurs due to the
local momentum of nuclear collision $K(r)=\{2Mm/\hbar^2[E-V_N(r)-
V_c(r)]\}^{1/2}$ where $Mm=A_pA_tm/(A_p+A_t)$ is the reduced mass. The
effective $v_{NN}$ potential includes M3Y force and also the factor
$F(\rho)=C[1+\alpha\exp(-\beta\rho)-\gamma\rho]$ which depends on densities
$\rho= \rho_p + \rho_t$, and the factor $(1-0.003\,E/A_p)$ that corrects
the dependence upon energy. All the parameters are well
established from many applications to the heavy-ion scattering data.
(Details one can see, e.g., in \cite{KS}). However, one should remind that
in such a semi-microscopic model the imaginary potential is introduced in
a phenomenological way, while the HEA approach uses the microscopic
model for both terms of the potential, where each of them depends on energy
as well.

Comparison of (\ref{eq5}) with (\ref{eq7}) ensures that the HEA potential
$V^H$ is only compose the direct part of the full potential while the real DF
potential $V^{DF}$ consists of two terms, direct and exchange one. Moreover,
the latter term takes into account not only the Pauli-blocking effect but
also effects of the knock-on reactions on the elastic scattering potential.

These two kinds of potentials $V^{DF}$ and $V^H$, $W^H$ having slightly
different slopes in asymptotics are employed to construct the total
microscopic potentials. In addition, one should bear in mind that at high
energies only the outer region of colliding nuclei play the essential role,
while the
exchange effects reveal themselves mainly in the central region. At the same
time, we pay attention to the result of \cite{W} that at high energies the
nucleons removal reactions contribute mostly to the absorption potential.
And this means that one-particle nuclear density distributions take part in
formation of respective matrix elements of an imaginary potential in the same
way as they participate in constructions of the real part of optical
potentials. Thus one can assume that tails of the real and imaginary parts
are almost of the same form, and thus we can utilize the pattern $V^{DF}$ to
build imaginary potential, too, at least in its outer region. As a result
we test the following forms of two-parameter potentials:
%[8]
\begin{equation}\label{eq8}
U^A_{\scr{opt}}~ = ~N^A_{r}\,V^H ~ + ~ iN^A_{im}\,W^H \qquad \,
\end{equation}
%[9]
\begin{equation}\label{eq9}
U^B_{\scr{opt}} ~ = ~ N^B_{r}\,V^{DF} ~ + ~ iN^B_{im}\,W^H \quad
\end{equation}
%[10]
\begin{equation}\label{eq10}
U^C_{\scr{opt}} ~ = ~ N^C_{r}\,V^{DF} ~ + ~ iN^C_{im}\,V^{DF}
\end{equation}
It is known that for heavy-ion scattering at comparably high energy only
potential tails determine the shape of differential cross-sections of elastic
scattering because of the strong absorption at small distances. Therefore,
in phenomenological consideration, one can limit himself by only the
4-parameter potentials like $U_{opt}=V_0\exp(-r/a_r) + i W_0\exp(-r/a_{im}$.
In our case we use the microscopic models for real and imaginary patterns
of optical potentials (\ref{eq8})-(\ref{eq10}), and by fitting two weight
factors $N_r$ and $N_{im}$ we can, in fact, change the strength and shift
the potential tails in the surface region. In practice the fit of
phenomenological potentials at $E\sim 100$ Mev/nucleon shows that the range
from $R_{in}$ to $\infty$ determines the main form of the differential
cross-sections, and $R_{in}$ is the radius where $V(R_{in})=-50$ Mev. So,
one can characterize the adjusted potentials
(\ref{eq8})-(\ref{eq10}) by volume integrals in the range from $R_{in}$ to
$\infty$. (Note, that in \cite{Simbel} integration runs from the rms radius
of a potential to $\infty$). So, we define them as follows
%[11]
\begin{equation}\label{eq11}
J^{out}_r={4\pi\over A_pA_t}\,\int\limits_{R_{in}}^\infty r^2dr\, V(r),
\qquad\quad
J^{out}_{im}={4\pi\over A_pA_t}\,\int\limits_{R_{in}}^\infty r^2dr\, W(r),
\end{equation}
and in futher will use them for comparisons of tested potentials.

%\large
\section {Results of calculations and conclusions}

We calculate the ratio of differential cross-sections $d\sigma/d\Omega=
|f(q)|^2$ to the Rutherford one
%[12]
\begin{equation}\label{eq12}
{d\sigma_R\over d\Omega}\,= \, \el ({Z_pZ_te^2\over \hbar v}\er )^2\,
{1\over 4k^2}\, {1\over\sin^4(\vartheta/2)},
\end{equation}
using the scattering amplitude in the framework of the
high-energy approximation:
%[13]
\begin{equation}\label{eq13}
f(q) = ik \int_0^\infty db b\, J_0 (qb)\Bigl [1-{\dis e}^
{\dis i\Phi_N(b)+i\Phi_C(b)}\Bigr ].
\end{equation}
\\
This expression is valid for $E\gg |U|$ and at small scattering angles
$\vartheta < \sqrt{2/kR}$ with $R$, the nucleus-nucleus interaction radius,
say, $R\sim {R_p+R_t}$. Here $q=2k\sin(\vartheta/2)$ is the transfer momentum.
The Coulomb phase $\Phi_C(b)$ is taken in analytic form for the potential
caused by the uniform charge density distribution. The nuclear phase
$\Phi_N(b)$ is calculated for the microscopic HEA- and DF-models as discussed
in the preceding section. The trajectory distortion is made, in nuclear phase,
by exchanging the impact parameter $b$ by $b_c=\bar a+\sqrt{{\bar a}^2+b^2}$,
the distance of closest approach in a Coulomb field, where $\bar a=Z_pZ_te^2/
2E$. Details of calculations of (\ref{eq12}) one can find in \cite{LZ2}. In
addition, we take into account effects of relativization by introducing
the respective relativistic velocity $v$ in the phase (\ref{eq3}) and in
the c.m. momentum $k$ in the amplitude (\ref{eq13}), and also in (\ref{eq11})
as follows:
%[14]
\begin{equation}\label{eq14}
\hbar v\,= \,197.327\,{\sqrt{E_l(E_l+2A_pm)}\over E_l+A_pm} \quad
(in ~~MeV\,fm),
\end{equation}
%[15]
\begin{equation}\label{eq15}
k\,=\,{1\over 197.327}\,\frac{A_t\sqrt{E_l(E_l+2A_pm)}}
{\sqrt{(A_p+A_t)^2\,+\,2A_pE_l/m}} \quad (in~~fm^{-1}).
\end{equation}
\\
Here $E_l$ (in MeV) -  kinetic energy of the projectile nucleus in
lab. system, and $m$=931.494 (in MeV) is the unified atomic mass unit.

Below we present our calculations of $d\sigma/d\sigma_R$ for scattering
of heavy ions ${^{17}O}$ on nuclei ${^{60}Ni},~{^{90}Zr},~{^{90}Sn},~
{^{208}Pb}$ at energy $E_l$=1435 MeV and compare them with the corresponding
experimental data from \cite{Lig}. The pattern potentials $V^H,~W^H,~V^{DF}$
were computed with help of eqs.(\ref{eq4})-(\ref{eq7}), and for this aim
we use the point density distributions of nuclei $\ro(r)$ from \cite{PP}
and \cite{FaS} for ${^{17}O}$ and target nuclei, correspondingly. Also,
parametrization of $\bar\sigma_{NN}$ and $\bar\alpha_{NN}$ are taken from
\cite{CG} and \cite{S2}. The effective $v_{NN}$-forces of the kind CDM3Y6 are
taken from \cite{KSO}.

As an example of the ${^{17}O}+{^{208}Pb}$ scattering, in Fig.1a,b, the
microscopic potentials $V^H$ and $W^H$ are shown by dashed curves, while
$V^{DF}$ is done by dash-dotted curve. The phenomenological Woods-Saxon (WS)
potentials, the real and imaginary part, fitted with experimental data in
\cite{Lig} , are shown by solid lines. One sees that in the outer region
the slopes of the calculated and the fitted potentials bear a
great resemblance to each other. In Fig.1c, cross sections, computed in
the framework of HEA with a help of the $V^H+iW^H$ (dashes) and $V^{DF}+iW^H$
(dash-dots) potentials are shown and compared with experimental data
(circles). All curves have an exponential fall beyond the Coulomb rainbow
angle, and the cross-section for $V^H+iW^H$ potential is in qualitative
agreement with the experimental data. As to an applicability of the HEA
calculations, one should compare the HEA curve (solid) for the WS-potential
and the points of experimental cross-sections which occur in precisely
coincidence with numerical solutions of the Shroedinger equation for this
potential. In fact, the agreement between them takes place at small angles
$\vartheta < 5.5^\circ $ where HEA is valid by definition. Therefore, in
the further HEA-calculations for another target nuclei we
will adjust the constructed microscopic potentials (\ref{eq8})-(\ref{eq10})
with the so-called "pseudo-experimental data", namely, with the respective
solid HEA-curve for the fitted phenomenological WS-potentials. Doing so we
believe that the obtained microscopic potentials will explain the data in
the whole range of scattering angles if one then calculates cross-sections by
numerical solution of the Shroedinger equation.

\vspace{0.5cm}

{\samepage
\hspace*{0.3cm} {\Large {\bf Table 1.}~~{\it Optical potentials,
constructed from the\\
\hspace*{3.5cm} microscopic HEA- and DF-potentials}}

{\normalsize
\begin{center}
\begin{tabular}{|l|c|c|c|c|}
\hline \hline
                      % & & & &\\
&${^{17}O}+{^{60}Ni}$&${^{17}O}+{^{90}Zr}$&${^{17}O}+{^{120}Sn}$&${^{17}O}+{^{208}Pb}$ \\
                      % & & & & \\
\hline \hline
                      % & & & & \\
$U^A_{opt}$& ---    & ---  & --- & --- \\
                      % & & & & \\

$U^B_{opt}$& --- &$0.6V^{DF}+i\,0.9W^H$&$0.5V^{DF}+i\,0.9W^H$&$0.5V^{DF}+i\,1.3W^H$ \\
                      % & & & & \\
$U^C_{opt}$&$0.6V^{DF}+i\,0.6V^{DF}$&$0.6V^{DF}+i\,0.5V^{DF}$&$0.5V^{DF}+i\,0.5V^{DF}$
&$0.5V^{DF}+i\,0.8V^{DF}$ \\
                      % & & & & \\
\hline
\end{tabular}
\end{center}
}
}

\vspace{0.5cm}

In Figs.2-5, the microscopic potentials, shown in windows $a$ and $b$ by
dashed curves for the HEA potentials $V^H$ and $W^H$, and by dash-dotted lines
for $V^{DF}$, were obtained by adjusting the respective cross-sections to
those for the fitted phenomenological potentials (solid lines). They are
presented in the more sensitive domain of distances where potentials fall
down from the value -50 MeV. In general, the obtained potentials have almost
the same slopes as those for the fitted WS-potentials. The respective HEA
cross-sections are demonstrated in the $c$ windows of Figs.2-5, and they are
in acceptable agreement with cross-sections (solid lines) calculated in HEA
for the fitted WS-potentials. In Table 1 one can find the fitted normalizing
factors $N_r$ and $N_{im}$ of the real and absorptive parts of the microscopic
potentials.

Table 2 demonstrates the values of outer volumes of the tested optical
potentials. They have magnitudes to be closed together which slightly
decrease with increasing atomic number of target nuclei. At the same time
they are approximately twice less than integrals of the fitted WS potentials.
This is due to the longer tails of the WS potentials in asymptotics.

\vspace{0.5cm}

{\samepage
\hspace*{0.3cm} {\Large {\bf Table 2.}~~{\it The outer volumes of optical
potentials}}\\

{\normalsize
\begin{center}
\begin{tabular}{|c|c|c|c|c|}
\hline \hline
Potential   &  $^{17}$O$+^{60}$Ni  &  $^{17}$O$+^{90}$Zr  &  $^{17}$O$+^{120}$Sn  &  $^{17}$O$+^{208}$Pb   \\
\cline{2-5}
            & $J^{out}_r\qquad J^{out}_{im}$ & $J^{out}_r\qquad J^{out}_{im}$
            & $J^{out}_r\qquad J^{out}_{im}$ & $J^{out}_r\qquad J^{out}_{im}$\\
\hline \hline
$U^{WS}_{opt}$   & 73.1 $\quad$ 57.7 & 67.7 $\quad$ 49.5 & 59.7
\quad 53.7 & 50.3 \quad 47.3 \\
$U^B_{opt}$  & & 27.2 $\quad$ 19.4 & 24.0 \quad 17.5 & 18.2  $\quad$ 12.9 \\
$U^C_{opt}$  & 32.4 \quad 32.4 & 27.2 $\quad$ 27.3 & 24.0 \quad
24.0 & 18.2 $\quad$ 18.3 \\ \hline
\end{tabular}
\end{center}
}
}

\vspace{0.5cm}

We did not intend to achieve a perfect fit as experimentalists usually
demonstrate. However, we can conclude that our idea proves itself to
utilize the microscopic models as patterns for the further fit with the
experimental data. In fact, doing so we introduced no more than two
normilizing free parameters while the number of parameters in the
phenomenological Woods-Saxon optical potential is required at least twice
that number. Moreover, at high energies, one can be sure that
calculations of microscopic potentials in the outer region give true
predictions of their behavior in the very sensitive domain of heavy-ion
scattering.

\vspace*{1.5cm}
\begin{center}
{\Large ACKNOWLEDGMENTS}
\end{center}

The co-authors V.K.L. and B.S. are grateful to the Infeld-Bogoliubov Program
and the head of its nuclear theory branch Prof. W.Rybarska for support of
this work. One of us E.V.Z. thanks the Russian Foundation Basic Research
(project 03-01-00657).

\begin{figure}
\begin{center}
\epsfig{file=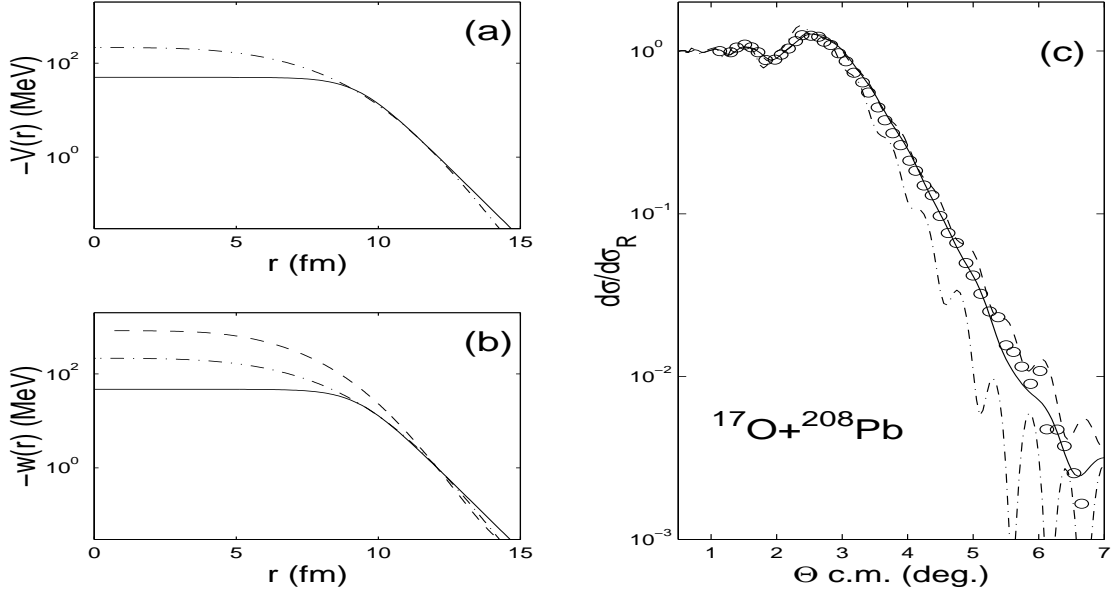,height=8cm,width=.9\linewidth}
\end{center}
\caption{Optical potentials $V^H(V^{DF})+iW^H$, calculated in the
microscopic approaches HEA(DF) for ${^{17}O}+{^{208}Pb}$ at
$E_l=1435$ MeV; ${\it a}$ - real part of potentials, ${\it b}$ -
imaginary part, ${\it c}$ - the ration of the respective elastic
to the Rutherford differential cross-sections. In windows ${\it
a}$ and ${\it b}$ dashed curves show potentials $V^H$ ¨ $W^H$,
dash-dotted lines - $V^{DF}$, solid curves
 - Woods-Saxon fitted potentials. In window ${\it c}$ showed cross-sections
calculated in the high-energy approximation (HEA): dashed curve -
for $V^H\,+\,iW^H$, dash-dotted - for $V^{DF}\,+\,iW^H$, solid
line - for the fitted WS-potential. Open circles - experimental
data from \cite{Lig}.}
\end{figure}

\begin{figure}
\begin{center}
\epsfig{file=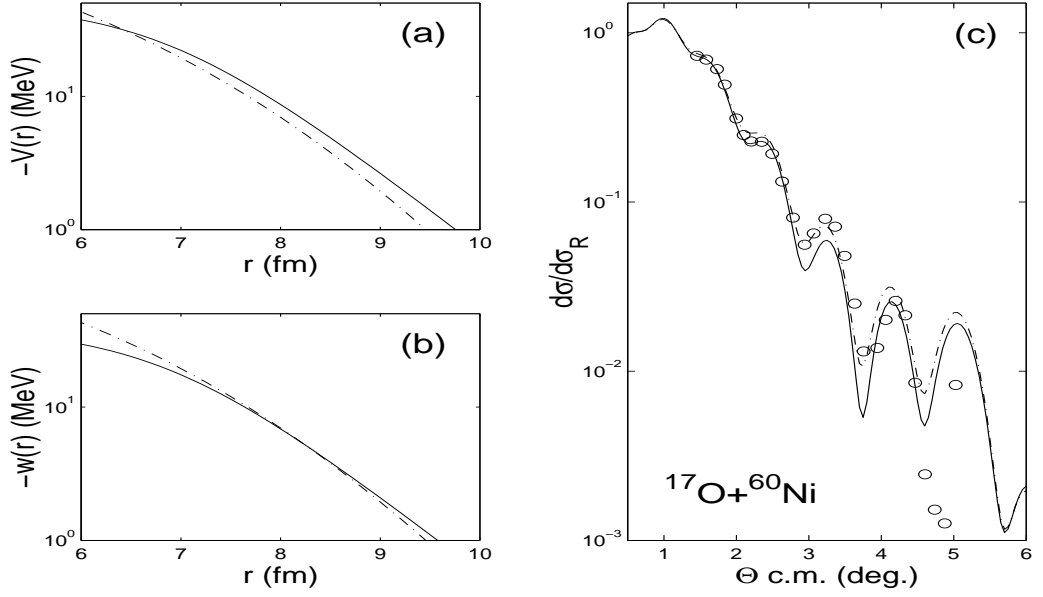,height=8cm,width=.9\linewidth}
\end{center}
\caption{The case of scattering ${^{17}O}+{^{60}Ni}$ at 1435 MeV.
In windows ${\it a}$ and ${\it b}$) are shown the microscopic
optical potential $U^C_{opt}$ constructed from the real $V^{DF}$
and imaginary $V^{DF}$ patterns multiplied by the fitted factors
$N_r$ and $N_{im}$ (dash-dotted curves); solid curves are the
fitted WS-potentials. The respective differential cross-sections
(window ${\it c}$), calculated in the HEA method. Solid curves are
cross-sections for the Woods-Saxon potential fitted with
experimental data \cite{Lig} ("pseudo-experimental data"),
dash-dotted curve is the cross-section for $U^C_{opt}$. The values
of $N_r$ and $N_{im}$ see in Table 1.}
\end{figure}

\begin{figure}
\begin{center}
\epsfig{file=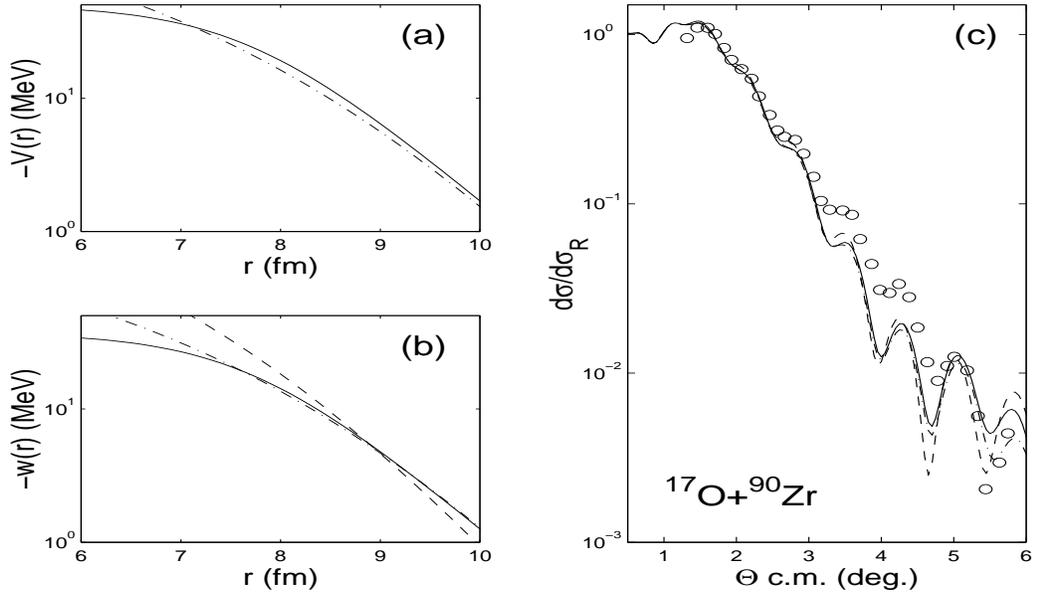,height=8cm,width=.9\linewidth}
\end{center}
\caption{The same as in Fig.2 but for scattering
${^{17}O}+{^{90}Zr}$. In addition, dotted curves correspond to the
$U^B_{opt}$ optical potential, composed from $V^{DF}$ and $W^H$
parts.}
\end{figure}

\begin{figure}
\begin{center}
\epsfig{file=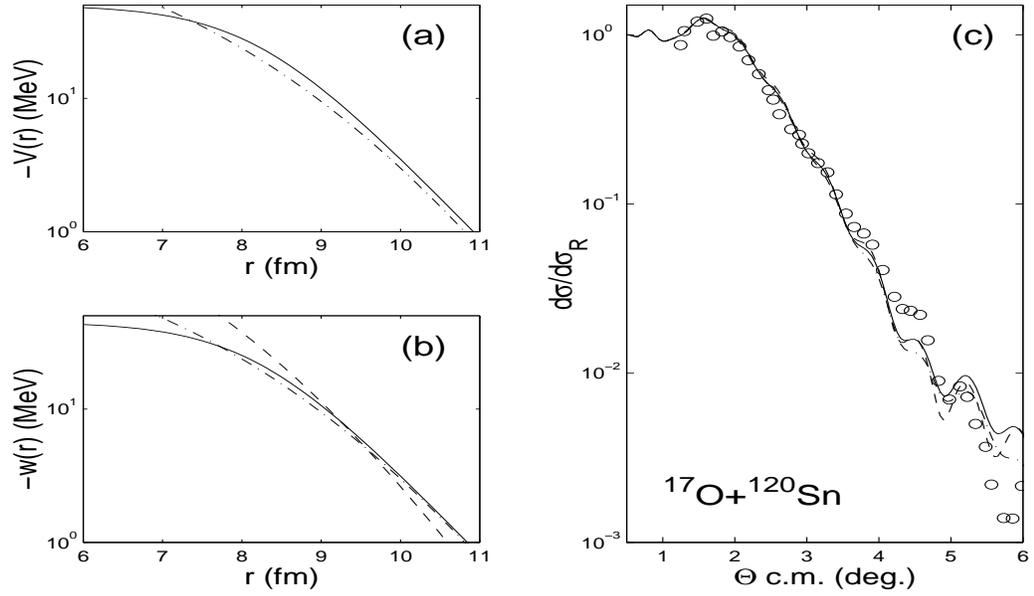,height=8cm,width=.9\linewidth}
\end{center}
\caption{The same as in Fig.3 but for scattering
${^{17}O}+{^{120}Sn}$.}
\end{figure}

\begin{figure}
\begin{center}
\epsfig{file=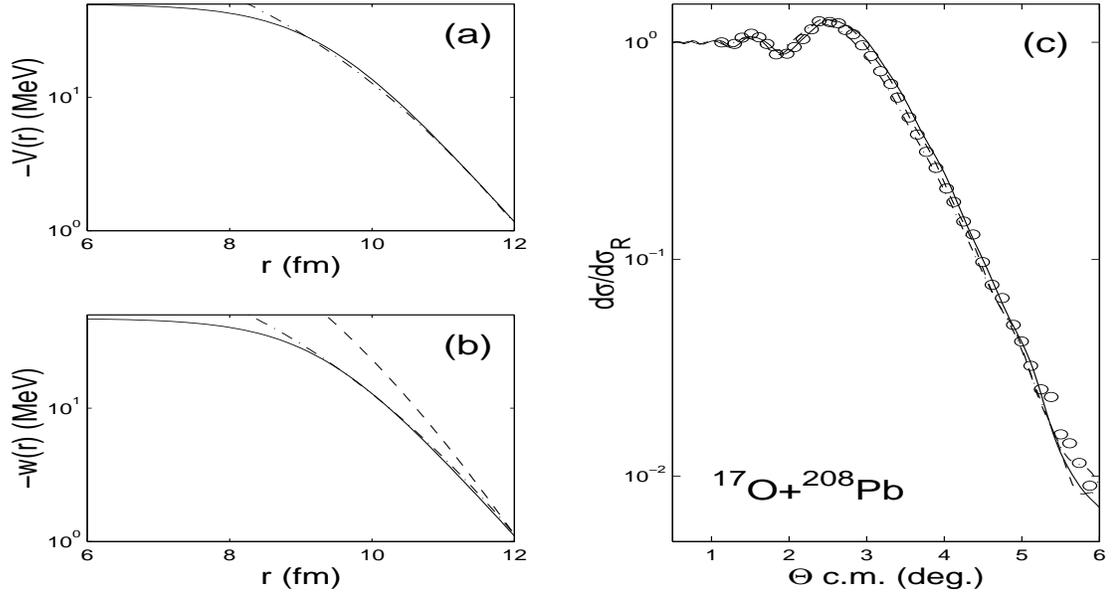,height=8cm,width=.9\linewidth}
\end{center}
\caption{The same as in Fig.3 but for scattering
${^{17}O}+{^{208}Pb}$. }
\end{figure}

\end{document}